\documentclass{article}
\usepackage{spconf,amsmath,graphicx}
\usepackage{hyperref}
\usepackage{amssymb}
\usepackage{booktabs}
\usepackage{multirow}
\usepackage{bbding}
\usepackage{xcolor}

\title{Magnitude-and-Phase-aware speech enhancement with parallel sequence modeling}
\name{Yuewei Zhang$^{1\ast}$ \qquad Huanbin Zou$^{2\ast}$ \qquad Jie Zhu$^{1\dagger}$\thanks{$^{\ast}$ Equally contribute to this work.} \thanks{$^{\dagger}$ Corresponding author (Email: zhujie@sjtu.edu.cn).}}
\address{$^{1}$ Department of Electronic Engineering, Shanghai Jiao Tong University, Shanghai, China \\
${^2}$ Tencent Video Cloud, Shanghai, China}
\copyrightnotice{979-8-3503-0689-7/23/\$31.00~\copyright2023 IEEE}
\begin{document}
%\ninept
%
\maketitle
\begin{abstract}
In speech enhancement (SE), phase estimation is important for perceptual quality, so many methods take clean speech’s complex short-time Fourier transform (STFT) spectrum or the complex ideal ratio mask (cIRM) as the learning target. To predict these complex targets, the common solution is to design a complex neural network, or use a real network to separately predict the real and imaginary parts of the target. But in this paper, we propose to use a real network to estimate the magnitude mask and normalized cIRM, which not only avoids the significant increase of the model complexity caused by complex networks, but also shows better performance than previous phase estimation methods. Meanwhile, we devise a parallel sequence modeling (PSM) block to improve the RNN block in the convolutional recurrent network (CRN)-based SE model. We name our method as \textbf{m}agnitude-and-\textbf{p}hase-aware and PSM-based \textbf{CRN} (\textbf{MPCRN}). The experimental results illustrate that our MPCRN has superior SE performance.
\end{abstract}
\section{Introduction}
\label{sec:intro}
In the real world, the clean speech is often contaminated by various types of environmental noise, leading to an noticeable decline in the perceptual quality and intelligibility of the speech. As a result, speech enhancement (SE) technique has become an important front-end speech signal processing step for numerous applications, such as voice communication, automatic speech recognition (ASR), and hearing aid devices. The primary objective of SE is to suppress the noise signal while preserving the valuable speech components. Traditional SE methods include spectral subtraction \cite{1163209}, wiener filtering \cite{1455809}, probabilistic modeling-based method \cite{NIPS2000_65699726}, etc. These methods heavily rely on specific prior assumptions and parameter settings, which limits their performance, particularly in the case of non-stationary noise pollution and low signal-to-noise ratio (SNR). In the past few years, many deep neural network (DNN)-based SE methods \cite{pascual2017segan,defossez20_interspeech,8547084,tan18_interspeech,hu20g_interspeech,9066933} have been proposed, demonstrating excellent performance compared to the traditional approaches.

While some works have attempted to directly enhance noisy speech in the time domain \cite{pascual2017segan,defossez20_interspeech}, a majority of recent studies have opted to tackle the SE task in the time-frequency (TF) domain \cite{8547084,tan18_interspeech,hu20g_interspeech,9066933}. TF domain methods tend to outperform the time-domain methods. One of the main advantages of TF domain methods is that their input contains more feature information, particularly the spectral features. Specifically, TF domain methods first transform the one-dimensional (1D) waveform into a two-dimensional (2D) spectrum using a short-time Fourier transform (STFT). The resulting 2D spectrum is then fed into a DNN for SE processing. Conventional TF domain methods \cite{8547084,tan18_interspeech} typically estimate the magnitude mask or directly the magnitude spectrum of the clean speech, and then reuse the original noisy speech's phase to reconstruct the enhanced speech. However, it has been demonstrated that phase recovery plays an important role in further improving the performance of SE \cite{PALIWAL2011465}. 

Nevertheless, predicting the clean speech's phase is challenging due to its lack of structural characteristics. Consequently, many works struggle to resolve the phase estimation problem in the SE task. For instance, \cite{Yin_Luo_Xiong_Zeng_2020} proposed a dual-branch network to separately estimate the magnitude and phase spectrum. In another approach, \cite{8682834} introduced a dual-branch network to simultaneously estimate the real and imaginary parts of STFT spectrum. Additionally, complex neural network has also been employed to directly predict the complex ideal ratio mask (cIRM) \cite{hu20g_interspeech,choi2018phaseaware}. However, these previous methods suffer from three main shortcomings. Firstly, both the dual-branch network and the complex network increase the parameter size and computational complexity of the SE model. The increase of model complexity is apparent and understandable in the case of the dual-branch network. As for the complex network, it replaces the ordinary real-valued convolutional layers, recurrent layers, and normalization layers in DNN with their complex-valued counterparts, which doubles the model size and quadruples computational operations. Therefore, this can be a disadvantage in terms of efficiency and practicality. Secondly, the approach that uses a real network to separately estimate the real and imaginary parts of the complex STFT spectrum or the cIRM has limited performance, since this method requires the network to learn the real and imaginary parts without prior knowledge \cite{hu20g_interspeech}. Thirdly, without the explicit magnitude and phase optimizations, implicitly enhancing the noisy speech in the complex STFT spectrum domain leads to the compensation problem \cite{9552504} between the magnitude and phase. As a result, this approach is susceptible to signal shifts in the time domain and adversely impacts the quality of the enhanced speech. %Thus, it is difficult to accurately predict the real and imaginary parts of the complex STFT spectrum or the cIRM by this approach.

In this paper, we propose a novel approach where the target magnitude and phase are estimated separately using the magnitude mask and normalized cIRM. These two new targets are easier for neural networks to predict. Meanwhile, the decoupling of the magnitude and phase estimation also improves the robustness of our method to temporal signal shifts. In addition, it has been observed that the real and imaginary parts of the audio’s STFT spectrum exhibit similar structural characteristics to its magnitude spectrum. Based on this insight, we employ a convolutional recurrent network (CRN)-based SE network with a single-branch network topology to simultaneously estimate the magnitude mask and normalized cIRM. Specifically, we utilize a parameter sharing strategy in which the output channel of the SE network's last decoder layer is set as 3. This configuration enables the simultaneous prediction of the magnitude mask, as well as the real and imaginary parts of the normalized cIRM. This strategy not only reduces the model size and computational complexity but also realizes a network regularization. Experimental comparisons with the previous phase-aware methods demonstrate the effectiveness and superiority of our proposed scheme.

Besides, we improve the recurrent module in the conventional CRN structure. The previous recurrent module only captures local and global dependencies along the time dimension, neglecting the speech’s spectral correlation in the frequency dimension. To address this limitation, we introduce a parallel sequence modeling (PSM) block as a replacement for the recurrent neural network (RNN) layer in the recurrent module. The PSM block adopts the parallel gated recurrent unit (GRU) and bidirectional gated recurrent unit (BiGRU) to perform sequence modeling along the input feature’s time and frequency dimensions, respectively. Subsequently, we employ a feature fusion network to integrate the two processed feature information and generate the fused result. The ablation study proves the performance benefits of our proposed PSM block.

In a nutshell, our contributions can be summarized as follows:
\begin{itemize}
    \item We propose to decouple the magnitude and phase estimation by simultaneously predicting the magnitude mask and normalized cIRM, which reduces the parameter size, computational complexity, and learning difficulty for the neural network, and finally yields an excellent SE performance.

    \item We introduce a PSM block to capture the sequential dynamics of speech features along both the time and frequency dimensions, which further improves the performance of the CRN-based SE model.
\end{itemize}

Combining the \textbf{m}agnitude-and-\textbf{p}hase-aware scheme and the proposed PSM block to improve the previous \textbf{CRN}-based model, we name our method as \textbf{MPCRN}.

The rest of the paper is organized as follows. Section \ref{sec:method} introduces the detailed architecture and principles of our proposed MPCRN. Section \ref{sec:exp} provides the experimental configurations, evaluation metrics, and experimental results. Section \ref{sec:conclusion} concludes the paper and discusses future research directions.

\begin{figure*}[htbp]
\centerline{\includegraphics[height=3.1in]{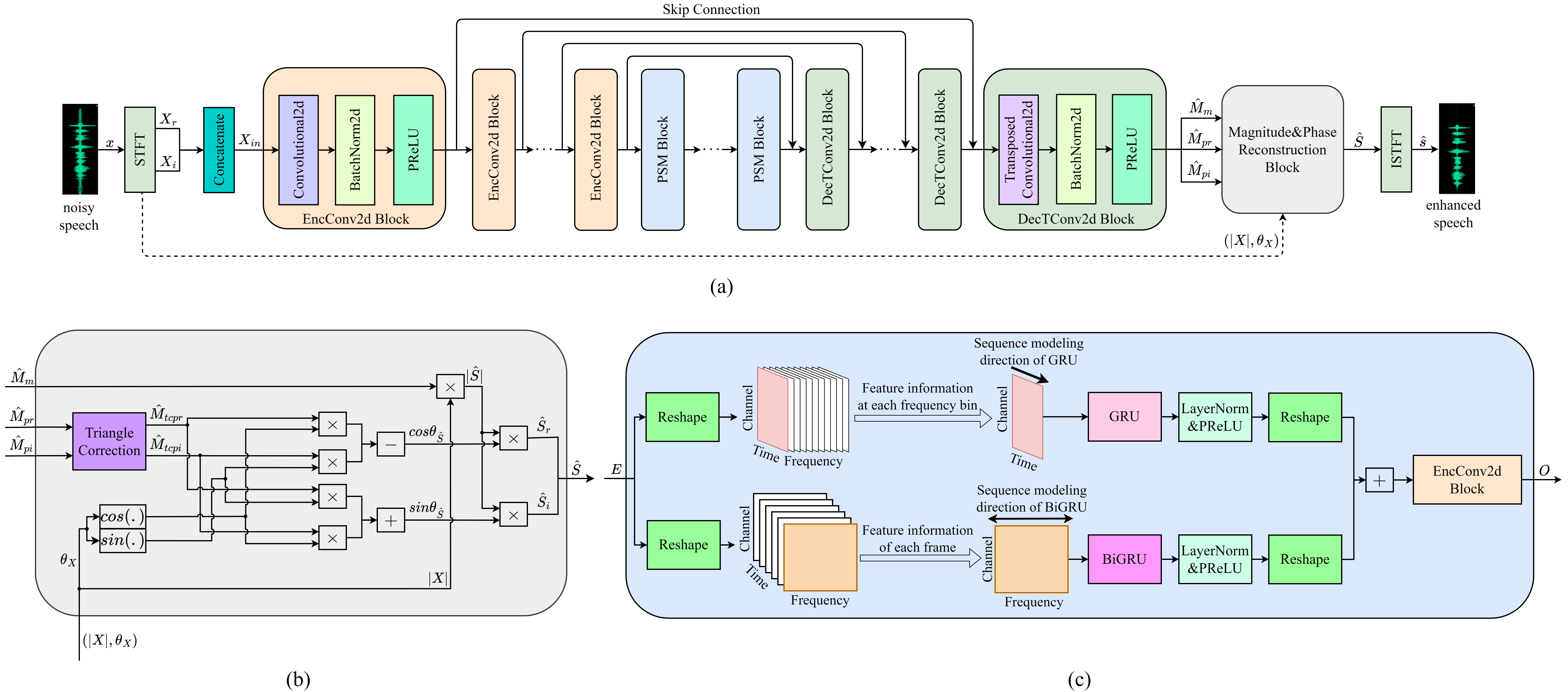}}
\caption{(a) Overall structure of MPCRN. (b) The diagram of magnitude\&phase reconstruction block. (c) The diagram of parallel sequence modeling (PSM) block.} \label{fig1}
\end{figure*}

\section{Method}
\label{sec:method}
\subsection{Signal Model}
In the time domain, the noisy speech $x(n)$ (where $n$ represents the discrete time index) can be formulated as
\begin{equation}
x(n)=s(n)+z(n)
\label{equation1}
\end{equation}
where $s(n)$ and $z(n)$ represent the clean speech and noise. Applying the STFT to both side of Eq. (\ref{equation1}), we obtain
\begin{equation}
X_{m,f}=S_{m,f}+Z_{m,f}
\label{equation2}
\end{equation}
where $X_{m,f}, S_{m,f}, Z_{m,f}\in\mathbb{C}$ are the STFT spectrum of noisy speech, clean speech, and additive noise. The $m$ and $f$ index the time frame and the frequency bin. In the following content, we omit the time and frequency indexes for brevity. In the Cartesian coordinates, Eq. (\ref{equation2}) can be written as
\begin{equation}
X_r+jX_i=(S_r+Z_r)+j(S_i+Z_i)
\label{equation3}
\end{equation}

In this work, we adopt the masking-based SE method, thus, the learning target of DNN can be expressed as
\begin{equation}
\begin{aligned}
M & = M_r+jM_i = \frac{S}{X} = \frac{S_r+jS_i}{X_r+jX_i} \\
%& = \frac{S}{X} \\
%& = \frac{S_r+jS_i}{X_r+jX_i}
%& = \frac{S_rX_r+S_iX_i+j(S_iX_r-S_rX_i)}{X_r^2+X_i^2}
\end{aligned}
\label{equation4}
\end{equation}
where $M\in\mathbb{C}$ is just the cIRM.

\subsection{Overall Network Architecture}
The overall network architecture of MPCRN is depicted in Fig. \ref{fig1}(a). Similar to CRN, MPCRN follows an encoder-decoder (ED) structure. The encoder is designed to extract high-level features from the input TF spectrum of the noisy speech, while the decoder aims to reconstruct the spectrum mask with the same resolution as the input spectrum. Additionally, there is a recurrent module between the encoder and decoder. The role of this recurrent module is to suppress the noise components within the extracted speech features and preserve the desired speech components.

In our work, we transform the input noisy speech $x$ by STFT and take the concatenated $X_{in}=Con(X_r,X_i)$ as the network input. The encoder consists of several 2D convolutional (EncConv2d) blocks, each composed of a 2D convolutional layer, a 2D batch normalization layer, and a PReLU layer. Following the encoder, the recurrent module comprises several of our proposed PSM blocks. These blocks replace the conventional RNN layers and provide a better modeling sequence capability. The details of our PSM block will be introduced in Section \ref{subsec:Parallel Sequence Modeling Block}. Next, the decoder, which has a symmetric structure to the encoder, includes several 2D transposed convolutional (DecTConv2d) blocks. Each DecTConv2d block contains a 2D transposed convolutional layer, a 2D batch normalization layer, and a PReLU layer. The decoder generates the predicted magnitude mask $\hat{M}_m$ and normalized cIRM $e^{j\theta_{\hat{M}}}=\hat{M}_{pr}+j\hat{M}_{pi}$. These predictions are then combined with the noisy magnitude $|X|$ and phase $\theta_{X}$ using a magnitude\&phase reconstruction block. The details of how this block performs magnitude and phase estimation for the enhanced speech will be presented in Section \ref{subsec:Magnitude and Phase Estimation}. Finally, the magnitude\&phase reconstruction block outputs the enhanced STFT spectrum $\hat{S}$, from which the enhanced speech $\hat{s}$ can be obtained by inverse STFT (ISTFT).

\subsection{Magnitude and Phase Estimation}
\label{subsec:Magnitude and Phase Estimation}
To achieve both magnitude and phase estimation for enhanced speech, previous masking-based SE methods always attempt to estimate the cIRM $M$ in Cartesian coordinates. In these methods, the prediction target is a complex value, and existing approaches either employ a complex network to directly estimate the cIRM or estimate the real part $M_r$ and imaginary part $M_i$ of the cIRM separately.

Different from the previous methods, we propose to decouple the magnitude and phase estimation and predict the cIRM in polar coordinates. The equivalent form of Eq. (\ref{equation2}) in polar coordinates is
\begin{equation}
|X|e^{j\theta_{X}}=|S|e^{j\theta_{S}}+|Z|e^{j\theta_{Z}}
\label{equation5}
\end{equation}
where $|\cdot|$ and $\theta_{(\cdot)}$ represent the magnitude and phase components.

Meanwhile, the cIRM $M$ in Eq. (\ref{equation4}) can also be rewritten as
\begin{equation}
M = |M|e^{j\theta_{M}}=\frac{S}{X}=\frac{|S|e^{j\theta_{S}}}{|X|e^{j\theta_{X}}}=\frac{|S|}{|X|}e^{j(\theta_{S}-\theta_{X})}
\label{equation6}
\end{equation}
where $|M|$ and $\theta_{M}$ are the magnitude and phase masks, and they can be obtained as
\begin{equation}
|M|=\frac{|S|}{|X|}
\label{equation7}
\end{equation}
\begin{equation}
e^{j\theta_M}=e^{j(\theta_S-\theta_X)}
\label{equation8}
\end{equation}

By this way, the prediction targets of the DNN are transformed into the magnitude mask $|M|$ and the phase mask $\theta_M$. Similar to the previous magnitude-only SE method \cite{8547084}, our MPCRN generates a bounded magnitude mask estimation $\hat{M}_m\in(0,1)$. However, directly estimating the phase mask is difficult due to the non-structural characteristics of speech phase. In our work, we choose to equally estimate $e^{j\theta_M}$, which corresponds to the normalized cIRM. Specifically, our MPCRN outputs two bounded tensors as $\hat{M}_{pr}\in(-1,1)$ and $\hat{M}_{pi}\in(-1,1)$, which are respectively the estimations for $cos(\theta_M)$ and $sin(\theta_M)$. Thus, the normalized cIRM is estimated as
\begin{equation}
e^{j\theta_{\hat{M}}}=\hat{M}_{pr}+j\hat{M}_{pi}
\label{equation9}
\end{equation}
where $\theta_{\hat{M}}$ is the estimated phase mask.

In order to predict the aforementioned $\hat{M}_m$, $\hat{M}_{pr}$ and $\hat{M}_{pi}$, we set the output channel of the 2D transposed convolutional layer in the last DecTConv2d block to 3. Then, we apply the Sigmoid function to the tensor of the first output channel, resulting in $\hat{M}_m$. Similarly, we apply the Tanh function to the tensors of the second and third output channels, yielding $\hat{M}_{pr}$ and $\hat{M}_{pi}$, respectively.

Combining the predicted masks (including the magnitude mask estimation $\hat{M}_m$ and the normalized cIRM estimation $\hat{M}_{pr}+j\hat{M}_{pi}$) and the noisy spectrum (including the noisy magnitude $|X|$ and the noisy phase $\theta_X$), we can obtain the enhanced spectrum $\hat{S}$. This process is achieved through the magnitude\&phase reconstruction block, as illustrated in Fig. \ref{fig1}(b). The detailed calculation process of this block is described below.

According to Eq. (\ref{equation7}), the enhanced magnitude $|\hat{S}|$ can be obtained as
\begin{equation}
|\hat{S}|=\hat{M}_m\cdot|X|
\label{equation10}
\end{equation}

The enhanced phase is derived from the normalized cIRM and the noisy phase. Firstly, since the $\hat{M}_{pr}$ and $\hat{M}_{pi}$ are respectively the estimations for $cos(\theta_M)$ and $sin(\theta_M)$, they should satisfy the condition $\hat{M}_{pr}^2+\hat{M}_{pi}^2=1$. But in practice, it is hard to guarantee this condition, because $\hat{M}_{pr}$ and $\hat{M}_{pi}$ are the outputs of DNN. Therefore, we modify $\hat{M}_{pr}$ and $\hat{M}_{pi}$ using triangle correction, i.e.,
\begin{equation}
\hat{M}_{tcpr}=\frac{\hat{M}_{pr}}{\sqrt{{\hat{M}_{pr}}^2+{\hat{M}_{pi}}^2}}
\label{equation11}
\end{equation}

\begin{equation}
\hat{M}_{tcpi}=\frac{\hat{M}_{pi}}{\sqrt{{\hat{M}_{pr}}^2+{\hat{M}_{pi}}^2}}
\label{equation12}
\end{equation}

Thus, the estimated cIRM in Eq. (\ref{equation9}) should also be modified as
\begin{equation}
e^{j\theta_{\hat{M}}}=\hat{M}_{tcpr}+j\hat{M}_{tcpi}
\label{equation13}
\end{equation}

According to Eq. (\ref{equation8}), the enhanced phase $\theta_{\hat{S}}$ can be obtained as
\begin{equation}
\begin{aligned}
e^{j\theta_{\hat{S}}} & = e^{j\theta_{\hat{M}}}\cdot{e^{j\theta_{X}}} \\
& = (\hat{M}_{tcpr}+j\hat{M}_{tcpi})\cdot{(cos(\theta_X)+jsin(\theta_X))}
\end{aligned}
\label{equation14}
\end{equation}

Extracting the real and imaginary terms on both sides of Eq. (\ref{equation14}) yields
\begin{equation}
cos(\theta_{\hat{S}})=\hat{M}_{tcpr}\cdot{cos(\theta_X)}-\hat{M}_{tcpi}\cdot{sin(\theta_X)}
\label{equation15}
\end{equation}
\begin{equation}
sin(\theta_{\hat{S}})=\hat{M}_{tcpr}\cdot{sin(\theta_X)}+\hat{M}_{tcpi}\cdot{cos(\theta_X)}
\label{equation16}
\end{equation}

Based on the results of Eq. (\ref{equation10}) and Eq. (\ref{equation15}, \ref{equation16}), the real and imaginary parts of the enhanced spectrum can be derived as
\begin{equation}
\hat{S}_r=|\hat{S}|\cdot{cos(\theta_{\hat{S}})}
\label{equation17}
\end{equation}
\begin{equation}
\hat{S}_i=|\hat{S}|\cdot{sin(\theta_{\hat{S}})}
\label{equation18}
\end{equation}

\subsection{Parallel Sequence Modeling Block}
\label{subsec:Parallel Sequence Modeling Block}
The conventional CRN architecture incorporates multiple RNN layers between the encoder and decoder to capture temporal correlations in the speech features and serve for noise reduction. However, modeling the local and global spectral dependencies among different frequency bins in speech are also crucial for SE. Therefore, we introduce the sequence modeling along the frequency dimension to further enhance the SE performance. To achieve this, we design a PSM block to replace the previous RNN layer in the CRN architecture.

The details of our proposed PSM block is illustrated in Fig. \ref{fig1}(c). This block consists of two main parts: a dual-branch sequence modeling network and a feature fusion network. Once the input feature $E$ is fed into the PSM block, the dual-branch network captures the sequential context of the input feature along both the time and frequency dimensions simultaneously.

For temporal sequence modeling, we reshape the input feature $E$ to ensure that the subsequent GRU layer can model the correlation among different speech frames. The GRU layer is followed by a layer normalization and a PReLU function, which is beneficial to the generalization and representation capability of the network. The result after PReLU function is reshaped again, so that the processed feature after temporal sequence modeling has the same dimensional order as the original input $E$.

For spectral sequence modeling, we reshape $E$ in another way, so as to use a BiGRU layer to model the frequency dynamics of the input feature at each frame. Since this sequence modeling process does not influence the causal inference, we adopt BiGRU instead of GRU, as it yields better performance. Same as the temporal branch, there is also a layer normalization, a PReLU function, and a reshape operation after the BiGRU, and their purposes are consistent to the temporal branch.

The feature fusion network receives the output features from the above two branches. It adds the two output features together and subsequently processes the sum through an EncConv2d block. This EncConv2d block contains a $1\times1$ convolutional layer, whose purpose is to further integrate the features from the previous two branches and adjust the number of channels in the output feature. Finally, the result after this EncConv2d block is just the output $O$ of our PSM block, and the output $O$ has the same shape as the input $E$.

\subsection{Loss Function}
Similar to previous works \cite{hu20g_interspeech,8910352}, we optimize our MPCRN by signal approximation (SA) \cite{7032183}, which aims to minimize the error between the enhanced speech and the clean speech. Thus, our loss function is formulated as
\begin{equation}
\mathcal{L}_{\text{mag}}=\left\|\sqrt{\hat{S}_r^2+\hat{S}_i^2}-\sqrt{S_r^2+S_i^2}\right\|_F^2
\label{equation19}
\end{equation}
\begin{equation}
\mathcal{L}_{\text{RI}}=\left\|\hat{S}_r-S_r\right\|_F^2+\left\|\hat{S}_i-S_i\right\|_F^2
\label{equation20}
\end{equation}
\begin{equation}
\mathcal{L}=\alpha_1 \mathcal{L}_{\text{mag}}+\alpha_2 \mathcal{L}_{\text{RI}}
\label{equation21}
\end{equation}
where $\mathcal{L}_{\text{mag}}$ and $\mathcal{L}_{\text{RI}}$ are the magnitude spectral loss and the complex spectral loss. $\left\|\cdot\right\|_F^2$ denotes the mean square error (MSE) loss. In Eq. (\ref{equation21}), the total loss $\mathcal{L}$ is the weighted sum of $\mathcal{L}_{\text{mag}}$ and $\mathcal{L}_{\text{RI}}$. In our work, we set the weights of the two loss items as $\alpha_1=\alpha_2=1$.

\section{Experiments}
\label{sec:exp}
\subsection{Dataset}
We adopt the widely used VoiceBank+DEMAND dataset \cite{valentini2016investigating} to evaluate our method. This dataset includes 11,572 clean-noisy utterance pairs for training and another 824 clean-noisy utterance pairs for testing. For the training set, the clean audios are selected from 28 speakers' recordings of the Voice Bank corpus \cite{6709856}. These clean audios are mixed with noise (including 2 types of artificially generated noise and 8 types of noise recordings from the Demand database \cite{10.1121/1.4806631}) at the mixed SNRs of \{0dB,5dB,10dB,15dB\}. For the test set, the clean audios are from 2 unseen speakers' recordings of the Voice Bank corpus, and they are mixed with 5 unseen types of noise from the Demand database at the mixed SNRs of \{2.5dB,7.5dB,12.5dB,17.5dB\}. All utterances are resampled to 16KHz. During the model training process, all utterances are chunked to 3 seconds.

\subsection{Experimental Setup}
We employ a Hamming window to implement the STFT in our experiment. The window length and hop size are set as 32ms and 8ms, resulting in a 75\% overlap between consecutive frames. We use a 512-point FFT to compute the STFT spectrum, so the frequency dimension of the obtained STFT spectrum is 257.

In our MPCRN architecture, the encoder, recurrent module, and decoder respectively include five EncConv2d blocks, three PSM blocks, and five DecTConv2d blocks. For all the convolutional and transposed convolutional layers in the encoder and decoder, we set the kernel size as (5,2) in the frequency and time dimensions, and the stride is (2,1). The output channel of each convolutional layer in the encoder is \{16,32,64,128,256\}, while the output channel of each transposed convolutional layer is \{128,64,32,16,3\}. In each PSM block, the hidden units of the GRU layer and the BiGRU layer are the same. And the hidden units of the three PSM blocks are \{128,64,32\}, respectively. It is worth noting that we ensure causality in our MPCRN by using asymmetric zero-padding in all the convolutional and transposed convolutional layers. This enables our method to achieve real-time SE.

During the training stage, we utilize the RMSprop optimizer with an initial learning rate of 2e-4. The learning rate decays by 0.5 if the model performance does not improve for 6 consecutive epochs. We conduct a total of 100 epochs for model training, with a batch size of 16.

\subsection{Ablation Study}
The model performance is evaluated by the wide-band perceptual evaluation of speech quality (WB-PESQ) \cite{941023} and three MOS metrics (i.e., CSIG, CBAK, and COVL) \cite{4389058}.

We conduct an ablation study to demonstrate the effectiveness of our magnitude-and-phase-aware scheme and PSM block. The results of ablation study are presented in Table \ref{table1}. %In these experiments, we have modified the predicting target of MPCRN to cIRM, but we still adopt the SA approach to optimize the network. Meanwhile, similar to \cite{hu20g_interspeech}, we have attempted three different methods to calculate the estimated enhancement spectrum from the cIRM predicted by MPCRN. Specifically, denoting the predicted cIRM as $\hat{M}=\hat{M}_r+j\hat{M}_i$, we combine $\hat{M}$ with the input noisy spectrum $X=X_r+jX_i$ using three multiplicative patterns to obtain the enhanced spectrum $\hat{S}$ as followings.

\begin{table}[ht]
  \centering
  \caption{Ablation study on VoiceBank+DEMAND test set}
  \setlength{\tabcolsep}{1mm}{
    \begin{tabular}{lccccc}
    \toprule[2pt]
          & WB-PESQ & CSIG  & CBAK  & COVL \\
    \hline
    noisy & 1.97  & 3.35  & 2.44  & 2.63 \\
    \hline
    MPCRN-R & 2.80 & 4.00 & 3.41 & 3.39 \\
    MPCRN-C & 2.80 & 3.97 & 3.42 & 3.38 \\
    MPCRN-E & 2.86 & 4.08 & 3.45 & 3.47 \\
    \hline
    MPCRN-w/o-PSM & 2.81 & 4.02 & 3.43 & 3.41 \\
    \hline
    MPCRN & \textbf{2.96}  & \textbf{4.16} & \textbf{3.50} & \textbf{3.56} \\
    \bottomrule[2pt]
    \end{tabular}}%
  \label{table1}%
\end{table}%

\begin{table*}[ht]
  \centering
  \caption{Performance comparison with other advanced systems on VoiceBank+DEMAND test set under causal implementation. Unreported values of related work are indicated as "-".}
    \setlength{\tabcolsep}{3.5mm}{
    \begin{tabular}{lcccccccc}
    \toprule[2pt]
    Methods & Year & Input & WB-PESQ & CSIG  & CBAK  & COVL & Model Size (M)  \\
    \hline
    \hline
    noisy & - & - & 1.97  & 3.35  & 2.44  & 2.63 & - \\
    \hline
    RNNoise \cite{8547084} & 2018 & Magnitude & 2.29 & - & - & - & 0.06 \\
    NSNet2 \cite{9413580} & 2021 & Magnitude & 2.47 & 3.23 & 2.99 & 2.90 & 6.17 \\
    ERNN \cite{9054597} & 2020 & Magnitude & 2.54 & 3.74 & 2.65 & 3.13 & 0.79 \\
    CRN \cite{tan18_interspeech} & 2018 & Magnitude  & 2.56  & 3.51  & 2.98  & 3.02 & - \\
    DCCRN \cite{hu20g_interspeech} & 2020 & Complex  & 2.68 & 3.88 & 3.18 & 3.27 & 3.7 \\
    PercepNet \cite{valin20_interspeech} & 2020 & Magnitude & 2.73 & - & - & - & 8 \\
    DeepMMSE \cite{9066933} & 2020 & Magnitude & 2.77   & 4.14  & 3.32  & 3.46 & - \\
    %DCCRN+ \cite{lv2021dccrn} & 2021 & Complex &  \CheckmarkBold  & 3.3  & 2.84    & -     & -     & - \\
    LFSFNet \cite{chen22c_interspeech} & 2022 & Magnitude & 2.91    & -  & - & - & 3.1 \\
    DEMUCS \cite{defossez20_interspeech} & 2021 & Time & 2.93   & 4.22  & 3.25  & 3.52 & 128 \\
    GaGNet \cite{LI2022108499} & 2022 & Complex & 2.94   & \textbf{4.26} & 3.45  & \textbf{3.59} & 5.94 \\
    \hline
    MPCRN & 2023 & Complex & \textbf{2.96} & 4.16  & \textbf{3.50} & 3.56 & 2.09 \\
    \bottomrule[2pt]
    \end{tabular}}%
  \label{tabel2}%
\end{table*}%

Many existing methods \cite{hu20g_interspeech,choi2018phaseaware} estimate the enhanced spectrum $\hat{S}=\hat{S}_r+j\hat{S}_i$ in Cartesian coordinates. Typically, these methods utilize a DNN to predict the cIRM $\hat{M}=\hat{M}_r+j\hat{M}_i$. Subsequently, the cIRM is combined with the input noisy spectrum $X=X_r+jX_i$ to obtain the enhanced spectrum $\hat{S}$. Furthermore, as described in DCCRN \cite{hu20g_interspeech}, there are three multiplicative patterns to derive $\hat{S}$, which are named DCCRN-R, DCCRN-C, and DCCRN-E. To prove the superiority of our magnitude-and-phase-aware scheme over previous methods, we have also modified the predicting target of our MPCRN to cIRM, and calculated $\hat{S}$ using the same three patterns as in DCCRN. Correspondingly, we denote the three ablation experiments as MPCRN-R, MPCRN-C, and MPCRN-E, which can be expressed as followings.
\begin{itemize}
    \item MPCRN-R:
        \begin{equation}
        \hat{S}=(X_r\cdot{\hat{M}_r})+j(X_i\cdot{\hat{M}_i})
        \label{equation22}
        \end{equation}
    \item MPCRN-C:
        \begin{equation}
        \hat{S}=(X_r\cdot{\hat{M}_r}-X_i\cdot{\hat{M}_i})+j(X_r\cdot{\hat{M}_i}+X_i\cdot{\hat{M}_r})
        \label{equation23}
        \end{equation}
    \item MPCRN-E: 
        \begin{equation}
        \hat{S}=|X|\cdot{\sqrt{\hat{M}_r^2+\hat{M}_i^2}}\cdot{e^{\theta_X+arctan2(\hat{M}_i,\hat{M}_r)}}
        \label{equation24}
        \end{equation}
\end{itemize} 

In addition, we have also conducted the experiment without the PSM block, using only the ordinary GRU layer for sequence modeling. This configuration is denoted as MPCRN-w/o-PSM.

The evaluation results in Table \ref{table1} demonstrate that our MPCRN outperforms MPCRN-R, MPCRN-C, and MPCRN-E across all the evaluation metrics. This result confirms the advantages of our proposed magnitude-and-phase-aware scheme. In other words, taking the magnitude mask and normalized cIRM as the predicting targets yields superior performance compared to previous cIRM-based methods. Furthermore, we can also observe that MPCRN achieves better evaluation results than MPCRN-w/o-PSM, which verifies the benefit of our proposed PSM block.
%Analyzing the reason for this result, we believe that this is because the value range of the real and imaginary parts of cIRM is uncertain, so the existing cIRM-based methods do not use bounded activation functions to limit the predicted cIRM. However, this will increase the difficulty of network learning, leading to a degradation of model performance.

\subsection{Comparison with Other Advanced Systems}
We further compare our MPCRN with other advanced methods as shown in Table \ref{tabel2}. To ensure a fair model comparison, all the benchmarks are causal. Meanwhile, these benchmarks adopt various inputs and techniques, and they have demonstrated excellent performance during their respective evaluation periods. Thus, the comparison with these benchmarks will effectively highlight the superiority of our method. 

From the comparison results in Table \ref{tabel2}, we can find that our MPCRN outperforms the previous methods on most of the metrics. Notably, our method achieves the highest scores on WB-PESQ and CBAK, indicating its superiority in terms of perceptual speech quality and background noise reduction. Although DEMUCS \cite{defossez20_interspeech} performs better on CSIG, and GaGNet \cite{LI2022108499} performs better on CSIG and COVL, both of them have much more parameters than our MPCRN. Meanwhile, the computational operations of our MPCRN are 2.02 GMACs/s. We have also conducted an real-time factor test on Intel(R) Xeon(R) Platinum 8255C CPU@2.50GHz and the result is only 0.12, which is satisfactory. Thus, the low model complexity is another advantage of our MPCRN. 

In addition, since our MPCRN is causal, the inference delay is one frame duration, i.e., 32ms. Therefore, our model also satisfies the requirement for real-time denoising \cite{reddy20_interspeech}.

In a word, our MPCRN is a lightweight real-time SE model, and it demonstrates excellent performance compared to other advanced systems.

\section{Conclusion}
\label{sec:conclusion}
In this work, we have introduced MPCRN, a novel approach for real-time SE task. Our method addresses the phase estimation problem by representing the predicting target in the polar coordinates, namely the magnitude mask and normalized cIRM. The experimental results illustrate that our method outperforms the conventional phase-aware schemes. Additionally, our MPCRN model exhibits significantly fewer parameters compared to previous methods, as we adopt a CRN-based network to simultaneously estimate the magnitude mask and normalized cIRM. Furthermore, we have also proposed a PSM block to replace the RNN layer in the CRN architecture. This block effectively captures the sequential correlations of speech features in both time and frequency dimensions, which is demonstrated to be better than the ordinary RNN layer. In the future, our study should involve other tasks, such as speech dereverberation and speech separation.

\section{Acknowledge}
\label{sec:acknowledge}
This work was supported by the special funds of Shenzhen Science and Technology Innovation Commission under Grant No. CJGJZD20220517141400002.

\clearpage

% Below is an example of how to insert images. Delete the ``\vspace'' line,
% uncomment the preceding line ``\centerline...'' and replace ``imageX.ps''
% with a suitable PostScript file name.
% -------------------------------------------------------------------------
% \begin{figure}[htb]

% \begin{minipage}[b]{1.0\linewidth}
%   \centering
%   \centerline{\includegraphics[width=8.5cm]{image1}}
% %  \vspace{2.0cm}
%   \centerline{(a) Result 1}\medskip
% \end{minipage}
% %
% \begin{minipage}[b]{.48\linewidth}
%   \centering
%   \centerline{\includegraphics[width=4.0cm]{image3}}
% %  \vspace{1.5cm}
%   \centerline{(b) Results 3}\medskip
% \end{minipage}
% \hfill
% \begin{minipage}[b]{0.48\linewidth}
%   \centering
%   \centerline{\includegraphics[width=4.0cm]{image4}}
% %  \vspace{1.5cm}
%   \centerline{(c) Result 4}\medskip
% \end{minipage}
% %
% \caption{Example of placing a figure with experimental results.}
% \label{fig:res}
% %
% \end{figure}

%\vfill
%\pagebreak

\bibliographystyle{IEEEbib}
\bibliography{refs}

\end{document}